\documentclass{aa}  
\usepackage{graphicx}
\usepackage{txfonts}
\usepackage{natbib}
\def\am         {$^\prime$}
\def\as         {$^{\prime\prime}$}

\def\deg        {$^{\circ}$}
\def\ltsima{$\; \buildrel < \over \sim \;$} 
\def\simlt{\lower.5ex\hbox{\ltsima}} 
\def\gtsima{$\; \buildrel > \over \sim \;$} 
\def\simgt{\lower.5ex\hbox{\gtsima}} 
  
\begin{document}
   \title{Cold fronts in galaxy clusters}


   \author{Simona Ghizzardi
          \inst{1}
          \and
          Mariachiara Rossetti  \inst{1}
	  \and
           Silvano Molendi \inst{1}
          }

   \offprints{S. Ghizzardi \email{simona@iasf-milano.inaf.it}}

   \institute{INAF, Istituto di Astrofisica Spaziale e Fisica Cosmica, Via E. Bassini 15,
           I-20133 Milano, Italy }

   \date{}
\titlerunning{Cold fronts in galaxy clusters}
\abstract{Cold fronts have been observed in a large number of galaxy clusters.
Understanding their nature and origin 
is of primary importance for the investigation of the internal 
dynamics of clusters.}
{To gain insight on the nature of these features, we carry out a statistical 
investigation of their occurrence in 
a sample of galaxy clusters observed with  XMM-Newton and 
we correlate their presence with different cluster properties.
}
{We have selected a sample of 45 clusters starting 
from the B55 flux limited sample by Edge et al. (1990)
and performed a systematic search of cold fronts.}
{We find that a large fraction of clusters host at least one cold front.
Cold fronts are easily detected in all systems that are manifestly 
undergoing a merger event in the plane of the sky while the 
presence of such features in the remaining clusters is related to 
the presence of a steep entropy gradient, 
in agreement with theoretical expectations. Assuming that cold fronts 
in cool core clusters are triggered by minor merger events, 
we estimate  a minimum of 1/3 merging events per halo per Gyr.}
{}
 

   \maketitle
%

\section{Introduction}

The unprecedented angular resolution of the X-ray telescope {\it Chandra} led to the 
discovery of several new phenomena within various astrophysical systems.
One of these are cold fronts detected in galaxy clusters.
Initially observed in merging clusters,
the prototypes are found in A2142 \citep{Maxim:2000},  
A3667 \citep{Vikhlinin1:2001, Vikhlinin2:2001, Vikhlinin:2002} and 1E0657-56 \citep{Maxim:2002}. 
All these systems feature very sharp discontinuities in their X-ray images where
the drop of the surface brightness (and correspondingly of the gas density)
is accompanied by a jump in the gas temperature, with the denser 
region colder than the more rarefied region, unlike shock fronts.
For this reason, these features have 
been dubbed ``cold fronts'' \citep{Vikhlinin1:2001}. 
The density and the temperature discontinuities have similar 
amplitude so that pressure is approximatively continuous across the front.

Cold fronts have been  initially interpreted as the edge of the cool core 
of a merging 
substructure which has survived the merger and is rapidly moving through 
the ambient gas \citep{Maxim:2000}.

Cold fronts have successively been detected in the cores of some relaxed 
clusters (e.g. A1795: \citealp{Maxim:2001}; RX J1720.1+2638:
\citealp{Mazzotta:2001}; A496: \citealp{Dupke:2003}; 2A 0335+096: 
\citealp{Mazzotta:2003}) and to date
a large number of relaxed systems are known to host one.
Since the presence of cold fronts in cool cores provides evidence of gas 
motions and possibly of departures from hydrostatic equilibrium,
understanding the nature of such a widespread phenomenon is 
mandatory to characterize the dynamics of galaxy clusters.
High resolution hydrodynamical simulations are, at present, the main 
technique to investigate the mechanisms generating cold fronts. 
Indeed, cold front features could already be detected in simulations published 
prior to the launch of {\it Chandra}
\citep{Roettiger:1997, Roettiger:1998}.
After cold fronts discovery, several hydrodynamical simulations have been 
developed to model the effect of the 
ram-pressure stripping in a merger event and the formation 
of the cold front feature in merging clusters \citep{Heinz:2003, Nagai:2003,Mathis:2005}.
Several simulations have also been employed to understand the origin of cold fronts
in relaxed non-merging clusters \citep[e.g.][]{Churazov:2003, TH:2005, AM06}.

The emerging picture (\citealp{AM06}; see also \citealp{MM_review:2007} for a review)
is that cold fronts arise during major merging events 
through ram-pressure stripping mechanisms which induce the discontinuity 
among the merging dense subcluster and the less dense surrounding ICM.
In relaxed clusters, the cold fronts features are induced by
minor merger events which produce a disturbance on the gas in the core,
displace it from the center of the potential well and decouple it from the 
underlying dark matter through ram-pressure.
Subsequently, a sloshing mechanism sets in, generating cold fronts.
The necessary condition for triggering this mechanism is the 
presence of a steep entropy profile for the central gas which is generally 
fulfilled at the center of relaxed cool core clusters.

Cold fronts are at present observed in a large number of galaxy clusters. 
\citet{Maxim:2003} analyzed a sample of 37 relaxed clusters observed 
with Chandra showing that cold fronts are present in the 
majority of the cores of relaxed clusters. 
Recently, \citet{Owers:2009} characterized a sample of nine cold fronts with 
quantitative measurements of the thermodynamic discontinuities across the edges 
and associated the presence of a cold front with evidence of merger activity.\\
While many objects have been 
studied in detail to understand the nature of cold fronts, we still lack a systematic 
investigation of the characteristics of these phenomena and of their host 
clusters through a large sample. The aim of this paper is to perform a 
systematic search of cold fronts in a representative sample and to investigate the 
properties of their parent clusters. Such a study is necessary to inspect the 
nature and origin of cold fronts and eventually to test the reliability of the picture 
emerging from the simulations.

The sample is selected starting from the B55 flux limited sample by \citet{Edge:B55}.
We use for our analysis {\it XMM-Newton} data. In spite of its limited spatial resolution 
with respect to {\it Chandra}, {\it XMM-Newton} has the positive attribute of having 
a large field of view, allowing a significant coverage of most of the clusters. 
In most cases, the clusters are inside the 
EPIC field of view up to a radius  
$\simgt 0.3 r_{180}$, allowing the 
characterization of the main thermodynamical properties well beyond the core regions.
Additionally, the {\it XMM-Newton}
large collecting area allows a good statistics for a large number of objects. 

Among the several physical properties characterizing the 
intracluster medium, we focus our attention on entropy.
Entropy plays a key role in describing the thermodynamical state of the ICM,
its distribution is a signature of the thermodynamical history 
of the cluster and it is also intimately related to the non-gravitational processes 
which may have occurred 
\citep{Voit_entropy:2002, Voit_entropy:2005}.
Moreover, as previously stressed, simulations highlight how the steep gas entropy  
 profile is a necessary condition for the onset of the sloshing mechanism 
and therefore for the presence of cold fronts in cool core clusters \citep{AM06}.

The structure of the paper is the following.
In \S\ \ref{sec:CF_sample} we describe the sample of clusters that we have analyzed
and in \S\ \ref{sec:data_red} we provide details about the data reduction.
Then we describe (see \S\ \ref{sec:search_CF}) the algorithm used for the systematic 
search of cold fronts in the cluster sample.
We present our results about the occurrence and the origin of cold 
fronts in \S\ \ref{sec:occurr} and we discuss them in \S\ \ref{sec:disc}. We summarize 
our findings in \S\ \ref{sec:summary}.

We adopt a $\Lambda$CDM cosmology with $\Omega_{\rm{m}} = 
0.3$, $\Omega_\Lambda = 0.7$, and $H_0 = 70$ km s$^{-1}$ Mpc$^{-1}$.

\section{The sample}
\label{sec:CF_sample}

We use as a reference starting sample the flux limited sample 
by \cite{Edge:B55}. 
It includes 55 objects with 
flux $f_x > 1.7 \cdot 10^{-11}$ erg cm$^{-2}$ s$^{-1}$ in the 
2 - 10 keV energy band and is 90\% complete.
All the clusters are located at redshift $z <0.2$.

\begin{table*}
\caption{The list of the 45 clusters belonging to our selected sample.}            
\label{table:sample}     
\centering
\begin{tabular} {|l|c||l|c| }
\hline 
{\bf cluster name}  &  {\bf redshift} & {\bf cluster name} & {\bf redshift} \\
\hline
   Centaurus	&	0.0114	&	       A85	&	0.0551	\\

 A1060	&	0.0126	&	     A3532	&	0.0554	\\
A262	&	0.0163	&	A3667	&	0.0556	\\
      AWM7	&     	0.0172	&     	     A2319	&	0.0557	\\
    Perseus  	&	0.0176	&	   Cygnus A	&	0.0561	\\
     A1367	&	0.0220	&	     A2256	&	0.0581	\\
     A4038	&	0.0300	&	     A3266	&	0.0589	\\
     A2199	&	0.0302	&	     A3158	&	0.0597	\\
      A496	&	0.0329	&	     A1795	&	0.0625	\\
    2A0335	&	0.0349	&	      A399	&	0.0718	\\
     A2063	&	0.0349	&	     A2065	&	0.0726	\\
     A2052	&	0.0355	&	      A401	&	0.0737	\\
      A576	&	0.0389	&	     A3112	&	0.0750	\\
     A3571	&	0.0391	&	     A2029	&	0.0773	\\
      A119	&	0.0442	&	   A2255	&	0.0806	\\
     MKW3s	&	0.0450	&	     A1650	&	0.0838	\\
     A1644	&	0.0473	&	     A1651	&	0.0849	\\
     A4059	&	0.0475	&	     A2597	&	0.0852	\\
     A3558	&	0.0480	&	      A478	&	0.0881	\\
     A3562	&	0.0490	&	   PKS0745	&	0.1028	\\
    Triang. Aus.	&	0.0510	&	     A2204	&	0.1523	\\
    Hydra A	&	0.0538	&	     A1689	&	0.1832	\\
      A754	&	0.0542	&		&		\\
\hline
\end{tabular}

\end{table*}

Starting from this sample, we select all the clusters available in 
the {\it XMM-Newton} public archive. 
At the time of writing, Ophiuchus was not publicly available and 
the objects A2244, A644  had not been observed with {\it XMM-Newton}. For A754 we use 
the long observations we obtained from AO7 (P.I.: A. Leccardi).
Observations for 3C129, A2142, A2147, A1736, A3391 are highly affected by soft protons:
since their final good exposure time after cleaning procedures is generally below 5 
ks for MOS and $<0.5$ ks for {\it pn}, these clusters have been excised from the sample.
For clusters having more than one observation, we eliminate those observations with high 
soft protons contamination.
We also exclude Virgo and Coma: their extension does not allow a significant 
coverage within the EPIC field of view.

Our final sample is reduced to 45 objects.
In Table \ref{table:sample} we filed the list of clusters belonging to our sample.

\begin{figure}
   \centering
   \includegraphics[angle=90,width=9.5 truecm]
    {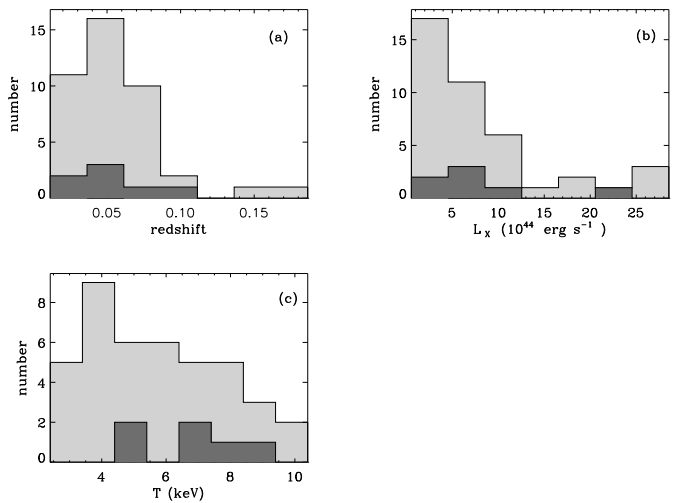}
   \caption{Distributions of redshifts (a), X-ray luminosities (b) and 
   temperatures (c) for the starting \citep{Edge:B55,Peres:1998} clusters sample 
   (light grey) and for the excised objects (dark grey). The rejection of these 
   objects does not introduce any obvious bias on the remaining subsample.}
   \label{fig: compl}%
\end{figure}

The excision of a number of clusters invalidates the completeness of the sample.
To verify if the final sample is representative of the cluster population, 
we inspect the distribution of the main cluster observables.
We built the histograms for redshift, X-ray luminosity and 
temperature (see Fig. \ref{fig: compl}) both for the 
starting sample (light grey) and for the excluded clusters (dark grey).
For X-ray luminosities ($L_X$) we refer to  \citet{Reiprich:2002}, while 
temperatures are taken from \citet{Peres:1998}.
The histograms show that excluded clusters do not introduce any obvious bias.
We conclude that, even if the final adopted sample is not complete,
it is representative of the cluster population.

\section{Data reduction}
\label{sec:data_red}

Observation Data Files (ODF) were retrieved from the {\it XMM-Newton}
archive and processed in a standard way with the Science Analysis System
(SAS) v6.1.

We apply the standard selection \verb|#XMMEA_EM| to the MOS event list 
(\verb|#XMMEA_EP| for {\it pn}) to automatically filter out artefact events.
The soft protons cleaning was performed using a double filtering process 
\citep[see][]{Leccardi_temp:2008, Pratt_arnaud_3sigma}.
The adoption of a threshold level and
the exclusion of light curve intervals  above the selected threshold 
allows the rejection of most flare events.
In practice, we extract the light curve in the 10 - 12 keV (10 -13 keV) energy band 
for MOS ({\it pn}) using 100 second bins.
We apply a threshold of 0.20 cts s$^{-1}$ for MOS and 0.50 cts s$^{-1}$ for {\it pn}
to generate  the filtered event file.
However softer flares
may exist such that their contribution above 10~keV is negligible.
To remove this flare contamination, we apply the 3$\sigma$ clipping method 
\citep[see][]{Marty_3sigma:2003}:
we extract a histogram of the light curve in the 2-5 keV band and fit this
histogram with a Gaussian distribution.
Since most flares have already been rejected in the previous step, the fit is
usually very good.
We then apply a threshold at the  $+3 \sigma$ level and
generate the filtered event file.
After soft proton cleaning, we filter the event file according to
\verb|FLAG| (\verb|FLAG|==0) and
\verb|PATTERN|  criteria (\verb|PATTERN|$\leq$12).

%

To systematically search surface brightness discontinuities, we build 
for each cluster, the EPIC flux map: MOS1 + MOS2 + {\it pn} ({\it pn} images are 
corrected for out of time events).
This flux image is computed in the 0.4 - 2 keV band following a 
procedure similar to 
the one described in \citet[see also \citealp{Rossetti_A3558:2007}]{Baldi_flux_map:2002}.
We sum up the
MOS1, MOS2 and {\it pn} source images to obtain the total source map, $S_{EPIC}$, and 
we compute EPIC 
source exposure map, $exp_{EPIC}$, by summing the source exposure maps of each
detector. The EPIC count rate image is then defined as  
$cr_{EPIC} =  S_{EPIC}/exp_{EPIC}$.
Count rates are then converted to flux through the total conversion factor $cf_{EPIC}$  
derived following the formula: 
$${{exp_{EPIC}}\over{cf_{EPIC}}} =
{exp_{MOS1}\over{cf_{MOS1}}}
+ {exp_{MOS2}\over {cf_{MOS2}}}
+ {exp_{pn} \over { cf_{pn}}}
$$
where $cf_{MOS1}, cf_{MOS2}, cf_{pn}$ and $exp_{MOS1}, exp_{MOS2},exp_{pn}$ 
are the conversion factors and the exposure maps of the three instruments.
The EPIC source flux image is obtained using the relation 
$fx_{EPIC} = cf_{EPIC} \cdot cr_{EPIC}$. 

To remove the quiescent particle induced background and the cosmic background component we need EPIC 
background flux images. 
We use a large collection of background data, such as long
observations of blank sky fields. We use 9 blank fields 
(for a total exposure time of $\sim$ 300 ks) selected by our
own group \citep{Leccardi_temp:2008}.
We compute the EPIC background flux image from the MOS1, MOS2
and {\it pn} background images and exposure maps using the same method applied to
the EPIC source flux image. By subtracting
the EPIC background flux image from the source flux image we derive
an EPIC net flux image in units of $10^{-15}$ erg cm$^{-2}$s$^{-1}$pixel$^{-1}$ 
(one pixel is $8.5 \times  8.5$ arcsec$^2$).
The net maps are used for the construction of the surface brightness profiles.

To search for cold fronts, we also need the temperature maps of all the clusters 
of our sample. We adopt 
a modified version of the ``adaptive binning + 
Broad Band Fitting'' algorithm described in 
\citet{Rossetti_A3558:2007}, where we have substituted the \citet{Cappellari:2003} 
adaptive binning algorithm with its modified version by \citet{Diehl:2006}.

\section{Systematic search of cold fronts}
\label{sec:search_CF}

\subsection{The detection algorithm}
\label{sub:detect}


In this section, we identify a suitable method to detect cold
fronts, well aware that any general selection criterion will have some 
limitations and can introduce spurious effects, so that some cold fronts may be
missed 
and some features may be classified as cold fronts although they are not.
We will address this point in detail in \S\ \ref{sub:notes}
and in  \S\ \ref{sub:occ_general}.

A cold front is characterized by the presence of a sharp decrease in the surface
brightness ($S\!B$) profile typically
accompanied by a rise of the gas temperature. 
We initially perform a systematic search of surface brightness 
discontinuities in the cluster sample and generate a list of candidate cold
fronts. 
Subsequently, we examine the gas temperature behavior across the discontinuity, in order 
to rule out the hypothesis that the detected discontinuity is a shock front.

\begin{figure}
\centering
  
\includegraphics[angle=0,width=8truecm]{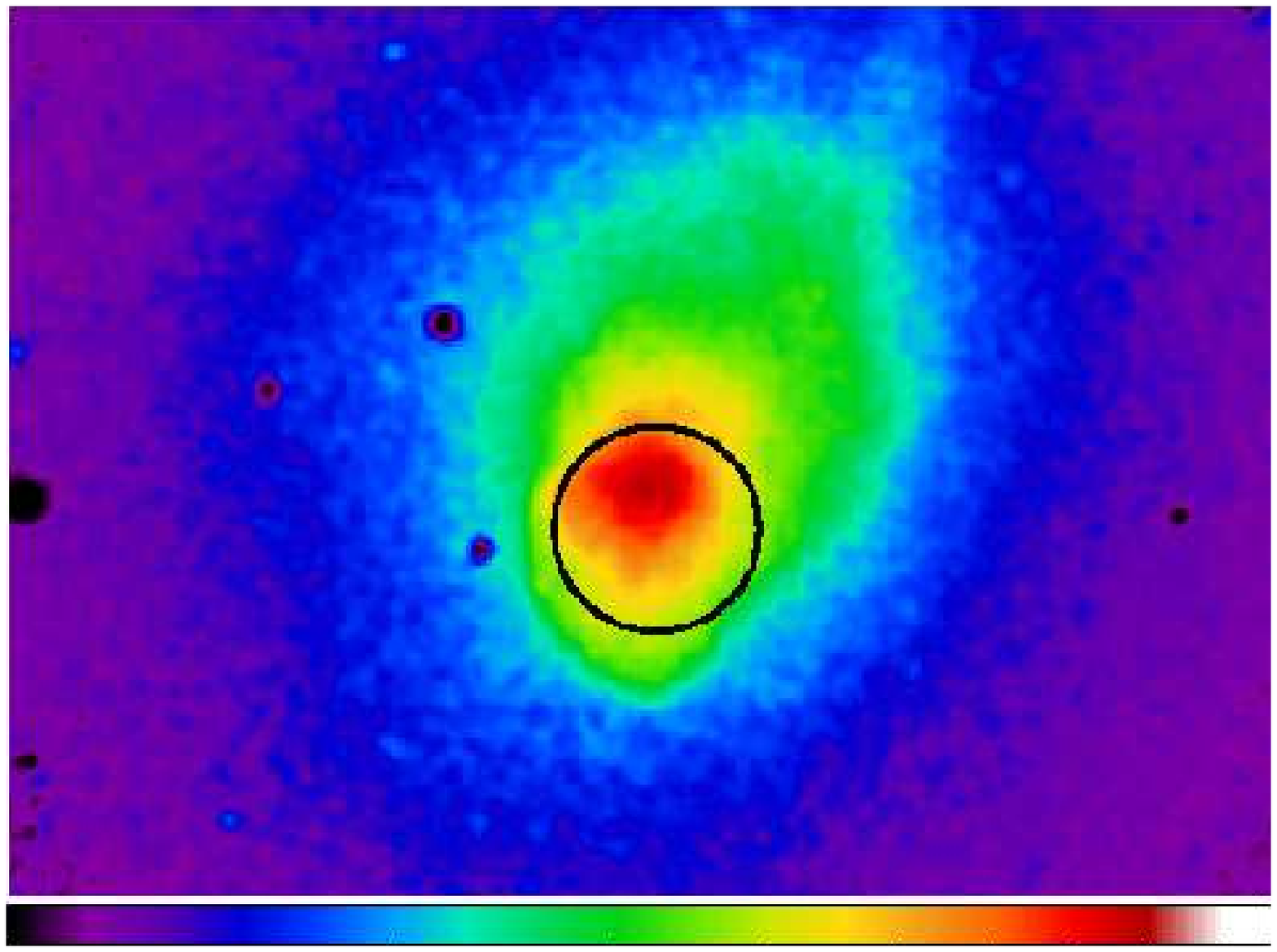}
\includegraphics[angle=0,width=8truecm]{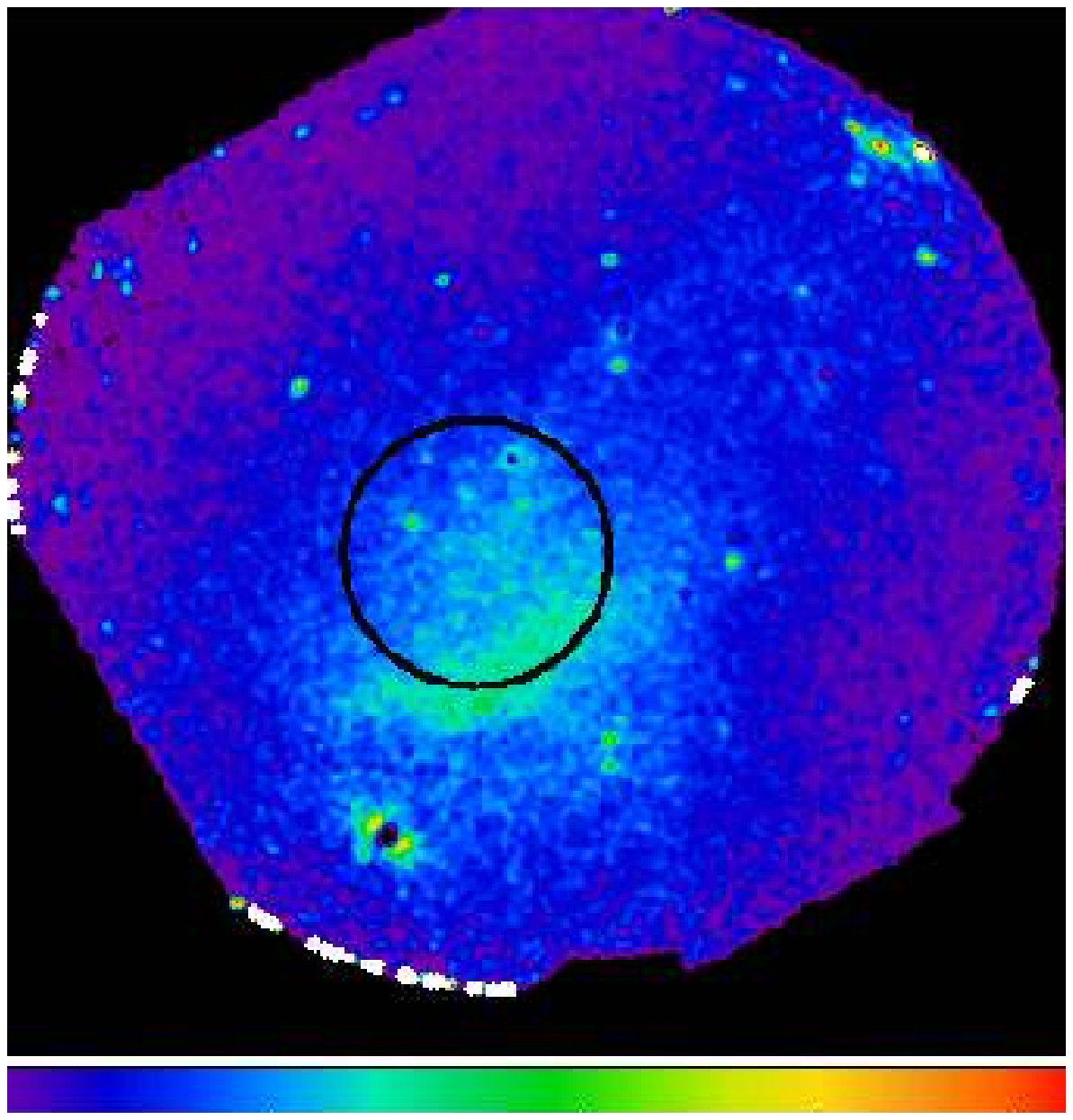}
 \caption{Surface brightness map for A2319 (top panel) and A1367 (bottom
panel). The maps have been smoothed for a better visual inspection. The black circles
show the position of the centers we chose for the radial profiles. They roughly match
the centers of curvature of the candidate discontinuity (SE-E for A2319 and S for
A1367) and do not match the X-ray emission peak. 
This choice  allows  a better characterization of the jump in the surface 
brightness profiles. Full resolution figures are available at: http://www.iasf-milano.inaf.it/$\sim$simona/pub/coldfronts/ghizzardi.pdf}
\label{fig:center}
\end{figure}

We developed an algorithm to perform the systematic search of surface brightness discontinuities.
We start from the EPIC flux maps (see \S\ \ref{sec:data_red}) that we have built 
for our sample and we divide
each cluster map in 30\deg\ wide sectors centered on the $S\!B$ peak. 
In most cases, we detect $S\!B$ discontinuities in several 
consecutive sectors, suggesting that most cold fronts have an angular extension 
significantly larger than 30\deg.
Consequently, unless the statistics is particularly low, 
fixed angular ranges can be used to find 
discontinuities. A possible bias on the detectability of cold 
fronts due to the width of the angular sectors will be discussed in \S\ \ref{sec:occurr}.
For clusters having a low
statistics or where the possible cold front is located near the cluster 
center (e.g. A262, A1795, A2199), we use
an ``ad hoc'' choice of the sectors (45\deg\ wide or even larger) to reveal the discontinuity in 
the surface brightness profiles.
We build the surface
brightness profile for each sector using the cluster X-ray emission peak as center.
Sometimes, for merging or irregular clusters,
a visual inspection of the images may suggest a different center,
better suited to detect a sharp decrease
of the surface brightness. In Fig. \ref{fig:center} we show the 
surface brightness images of A2319 and A1367 as an example. The black circles mark the centers
we adopted to build the profiles.

\begin{figure*}
   \centering
   \includegraphics[angle=90,width=15truecm]
{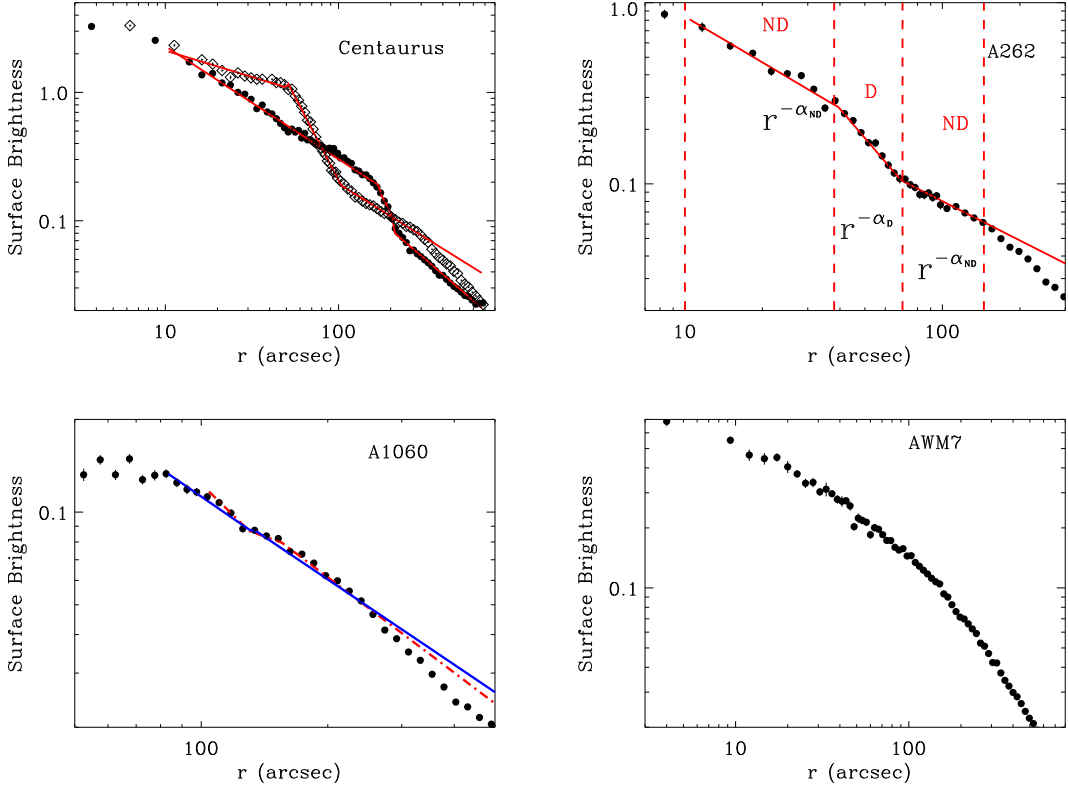}
   \caption{Surface brightness profiles for some sectors of four clusters of our sample. 
For the Centaurus cluster we plot two interesting sectors: SE (filled circles) and 
N-NW (open diamonds) of the cluster core. 
The figure shows that profiles may have different behaviors.
While discontinuities are apparent in some clusters (e.g.
Centaurus cluster), in others they are not as sharp. Some systems (e.g. AWM7)
have a regular profile. In upper right panel (A262) we plot the ranges used to
fit the profile with power laws. The flag D marks the discontinuity region and 
ND marks the adjacent (inner
and outer) regions. In all the panels the solid lines represent power law best fits 
(see text and Table 
\protect\ref{table:delta_s}  
for details). In these plots, surface brightness is given in 
$10^{-15}$ erg cm$^{-2}$ s$^{-1}$arcsec$^{-2}$ units.}
  \label{fig:discont}%
  \end{figure*}

For each cluster a set of profiles is obtained.
While for some clusters (e.g. Centaurus, A496, 2A0335+096) the presence of a
surface brightness discontinuity is apparent, in other clusters (e.g. A262) the 
surface brightness profile is not as sharp (see Fig. \ref{fig:discont}).
Projection effects and resolution limits smooth the profiles: the 
surface brightness discontinuity will appear as a steepening of the
profile in the radial range around the jump radius.  
In the approximation where profiles are described 
by power laws, the slopes measure the steepness of the profile.
We use the power law slopes 
to characterize the surface brightness discontinuities.
We identify for each cluster the possible discontinuity 
region with a visual inspection of the profile and of the image.
We mark this region with the letter D 
and we set $S\!B \propto r^{-\alpha_D}$ in the corresponding radial range 
(see upper right panel of Fig. \ref{fig:discont} as an example). 
We compare the slope we
find in this region with the slope obtained fitting the profile with the power law 
$S\!B \propto r^{-\alpha_{ND}}$ in the nearby 
(inner and/or outer) region or in other sectors of
the cluster where no irregularities in the surface brightness profiles are present.
The difference of the slopes $\Delta \alpha = \alpha_D - \alpha_{ND} $ quantifies 
the steepening of the profile.
We require that $\Delta \alpha \ge  0.4$ to classify a region as
discontinuous. The choice of this threshold relies on phenomenological considerations.
All the jumps we measured have $\Delta \alpha$ values well above 0.5, while for 
regions without discontinuities $\Delta \alpha$ 
values are below 0.2. 

Examples of different surface brightness profiles are reported in Fig.
\ref{fig:discont}.
 Centaurus cluster profiles (top left panel in the figure)
 steepen significantly between 50\as\ -- 100\as\ in the NE sector
and between 170\as\ -- 210\as\ in the W-NW sector. In A262 and in A1060
(top right panel and bottom left panel respectively) the discontinuities are
not as apparent as in Centaurus. 
Finally, the profile for AWM7 (bottom right panel) does not show
any irregularity. 
We fit all the profiles in the different radial ranges with power laws. In Table
\ref{table:delta_s} we 
report, for each region of the four clusters shown in Fig. \ref{fig:discont}, the
ranges used for the fits, the slopes of the best fits, and the associated $\Delta \alpha$
values.

\begin{table*}
\caption{Cold fronts candidates for the four clusters plotted in Fig. \ref{fig:discont}.
We report the cluster name, the position angle (measured anticlockwise from East) 
of the candidate cold front, the different radial range considered, 
the slopes of the power laws that we find in the given radial range and 
the $\Delta \alpha$ values for the discontinuities in the surface brightness profiles.
In the last column, a flag indicates which features are classified 
as discontinuities.}

\label{table:delta_s}      
\centering                          
\begin{tabular}{l c c c c c c }        
\hline\hline                 
cluster & position angle  &  radial range & $\alpha$ &  ranges for $\Delta \alpha$ &
$\Delta \alpha$ & discont. \\   
& (deg) & (arcsec) & & & & \\
\hline 
  Centaurus (internal)& [120,150] & 10 - 50  & 0.41 &  & & \\     
  Centaurus (external) & [120,150] & 100 - 240  & 0.84 &  & & \\ 
  Centaurus (possib disc) & [120,150] & 50 - 100  & 2.59 &
$\alpha_{[50-100]}-\alpha_{[10-50]}$ & 2.19 & $\surd$ \\
  Centaurus (possib disc) & [120,150] & 50 - 100  & 2.59 &
$\alpha_{[50-100]}-\alpha_{[100-240]}$ & 1.75 & $\surd$ \\
\hline                        
   Centaurus (internal)& [30,60] & 10 - 170  & 0.88 &  & & \\     
  Centaurus (external) & [30,60] & 210 - 600  & 1.16 &  & & \\ 
  Centaurus (possib disc) & [30,60] & 170 - 210  & 2.52 &
$\alpha_{[170-210]}-\alpha_{[210-600]}$ & 1.64 & $\surd$ \\
  Centaurus (possib disc) & [30,60] & 170 - 210  & 2.52 &
$\alpha_{[170-210]}-\alpha_{[10-170]}$ & 1.36 & $\surd$ \\
\hline 
  
   A262 (internal) & [-45,0] & 10 - 38  &  0.84 &  & &  \\
   A262 (external) & [-45,0] & 70 - 145  &  0.72 &  & &  \\
  A262 (possib disc) & [-45,0] & 38 - 70   &  1.66 &
$\alpha_{[38-70]}-\alpha_{[10-38]} $ & 0.82 & $\surd$ \\
   A262 (possib disc) & [-45,0] & 38 - 70  &  1.66 &
$\alpha_{[38-70]}-\alpha_{[70-145]}$ & 0.94 & $\surd$ \\
\hline 
A1060 (possib disc) & [120,150] & 80 - 135  & 0.94 &  & & \\
   A1060 (external) & [120,150] & 135 - 250 & 0.92 &
$\alpha_{[80-135]}-\alpha_{[135-250]}$ & 0.02 & $\times$ \\
 \hline
   AWM7 & [-90,-60] & 20 - 200  & 0.76 &  & 0.00 & $\times$ \\
\hline                                   
\end{tabular}
\end{table*}
For the Centaurus cluster, we consider two interesting sectors, SE and W-NW of
the cluster core. The slopes we find for the W-NW sector (specifically, 
120\deg\ -- 150\deg , where the angles are measured in an anticlockwise direction 
from East) are $0.41$, $2.59$
and $0.84$ for the radial ranges: 10\as\ -- 50\as, 50\as\ -- 100\as, 100\as\ --
240\as\ respectively. The $\Delta \alpha$
for the central region is 2.19 with respect to the innermost region 
and 1.75 with respect to the
outer one. This is obviously classified as a discontinuity and is a
candidate cold front.
A similar analysis allows to assess that there is a discontinuity in the 170\as\
--  210\as\ radial range in the 30\deg\ -- 60\deg\ sector. The quality of the
Centaurus
cluster profiles is extremely high thanks to its proximity and to the 
 long observations (170 ks in the public archive at
the time of writing) so that, even at the margin of the cold front located west
of the core, from -60\deg\ to 60\deg, the
discontinuity is still visible in the profile. The $\Delta \alpha$ we find 
for this case, 1.60,  is smaller than in the previous case, but still high.
 
In the top right panel, we show the profile for the SW-W region (-45\deg\ --
$\,\,$0\deg)
in A262, where a discontinuity at $\sim 60$\as, albeit not very sharp, is 
identifiable ($\Delta \alpha$ is slightly smaller than 1).
In A1060 (bottom left panel) the putative discontinuity is around 2\am. As
shown in Table \ref{table:delta_s}, the analysis of the slopes provide 
$\Delta \alpha = 0.02$. As a
consequence, this feature is not classified as discontinuity. However, a
different result 
could be obtained with a slightly different choice of the radial ranges used for
the fits.
If we restrict the range of the discontinuity region to  
110\as\ - 130\as\ (3 points for the fit) and we
choose 150\as\ - 250\as\  for the outer region (disregarding the 4 points
immediately after the discontinuity where the profile is flat) 
$\Delta \alpha$ increases to 0.55 and this feature could be considered as a
discontinuity. We believe that this choice of the radial range is rather
extreme. In addition, we note that the temperature profile
shows no variations in the same region.
As a general rule, if $\Delta \alpha$ satisfies the discontinuity condition 
only for an {\emph{ad hoc}} choice of the radial range, we exclude it from the list of 
candidate cold fronts.
Finally, in the bottom-right panel in Fig. \ref{fig:discont}, AWM7 shows a
regular behavior. In the figure we report one region, but AWM7 is
regular in all its sectors and all profiles are similar;
for this cluster we find no discontinuities.

\vskip0.3truecm
\noindent
The procedure described thus far detects surface brightness discontinuities and
provides a list of candidate cold fronts. To upgrade a discontinuity to a cold
front, we need to verify the behavior of the temperature profile 
across the surface brightness jump.
To this aim, we build the binned temperature maps (see \S\ \ref{sec:data_red} 
and \citealt{Rossetti_A3558:2007}) for all the clusters of our sample.  
From these maps we derive the temperature profiles
plotting all the bins whose barycentre is inside the
sector hosting the surface brightness discontinuity. In none of the candidate cold 
fronts we observe a sharp decrease in the temperature profile as would be expected for a shock front.
Almost all the surface brightness discontinuities that we find feature a sharp gas 
temperature rise. 
In some cases the temperature rises smoothly with no jumps. 
We remark that the cold front feature does not necessary require a temperature jump, 
since the thermal pressure of the gas inside the cold front is balanced by the sum 
of the thermal and ram pressures outside. This can be achieved also with a slow 
continuous rise of the temperature across the discontinuity.

\subsection{Notes on individual clusters}
\label{sub:notes}

In Tab. \ref{table:CF_sample}, we list the clusters
hosting one or more cold fronts. 
For each cluster, the table provides the center used to build the $S\!B$ profiles, 
the cold front, the radial and azimuthal  
position of the cold front and $\Delta \alpha$. The $\Delta \alpha$ reported 
for each cold front is the
mean value obtained averaging over the different sectors hosting the discontinuities.

As already remarked,
the procedure adopted to classify cold fronts can fail in finding the discontinuities 
or provide some doubtful cases. Hence, comments are required for some individual systems. 

\begin{itemize}
 
\item{{\bf A4059:} a visual inspection of the surface brightness map of A4059 
hints to a possible cold front in the SW sector $\sim 30$\as\ from the peak, 
but the discontinuity is barely visible in the profiles and we did not succeed in 
fitting it with power laws and deriving $\Delta \alpha$. We consider this 
as an unclassified case.}

\item{{\bf A85:}
This object has two possible cold fronts. The former lies in the NW sector 
$ \sim 80$\as\ 
from the X-ray peak. In this sector $\Delta \alpha = 0.75$, above our threshold. 
However, the sector of the cold front is very narrow (30\deg ) and 
no discontinuities are detected in the nearby sectors. Table \ref{table:delta_s}
shows that the cold fronts widths generally range from 60\deg\ to 120\deg .
Such a tiny extension for a cold front is unusual and we prefer to consider this as an 
unclassified case. Another cold front is present in a small subclump located 8\am\ 
south of the cluster 
and moving north towards the main structure \citep{Durret:98, Kempner:2002} . 
This is labeled as {\bf A85$^*$} in Table 
\ref{table:CF_sample} to distinguish it from the unclassified cold front of the 
main cluster.}

\item{{\bf A2052:} the analysis of the discontinuities in A2052 is complex because of 
the presence of the bright shells 
surrounding the X-ray cavities \citep{Blanton_A2052:2003, Blanton_A2052:2001}.
Remarkably, sharp decrements of the surface brightness profiles are detected just
 outside the rims, at distances of 
about 40\as -50\as\ (rims are at 10\as -30\as\ from the center).
It is difficult to establish whether such sharp drops are real discontinuities or if they 
are associated to the bright shells. Recent results obtained from a deep {\it Chandra} 
observation \citep{Blanton_A2052_ripples:2007}
show that some cavities and ripples are present in this cluster, similarly to what 
has been observed in Perseus clusters and M87 
\citep{Fabian_Perseus:2006, Fabian_Perseus:2003, Forman_M87:2007}. 
Some weak shocks may also be present. The presence of cold fronts is unclear and 
we consider this cluster as unclassified.}

\item{{\bf Hydra A:} in Hydra A, we detect a cold front in the 300\deg\ -- 330\deg\
sector  $\sim 50$\as\
from the core, inside the region of the weak shock which is at $\sim 200$\as\ 
\citep{McNamara_HyA_shock:2005,Nulsen_HyA_shock:2005}. Similarly to A85,
the sector of the cold front is narrow (30\deg ). The jump is located at the
bending of the south radio lobe 
\citep[visible in the Hydra A radio maps at 1.4 GHz; 
see for example][]{Lane_HyA_1.4GHz:2004}
towards east, near the SW cavity.  
In that region the temperature map shows several cold blobs and one 
of this produces the temperature rise  coincident with the surface brightness
steepening but no clear
front in the temperature map is present. 
The structure of Hydra A is very complex with a strong interaction 
between the radio lobes and the ICM gas 
(\citealp{Nulsen_HyA_radio:2002, McNamara_HyA_radio:2000}; see also 
\citealp{McNamara_review:2007}). 
The cluster exhibits a number of cavities in the central 
regions \citep{Wise_HyA_core:2007}. The discontinuity we find is probably a result 
of such a complicated morphology and  
likely it is not a cold front. Even if this region satisfies all the
required
conditions we consider this as an unclassified case.}

\end{itemize} 

Since the existence of a cold front in A2052, A4059 and Hydra A cannot be 
definitively established, we exclude these systems from the sample.

\begin{table*}
\caption{List of the cold fronts detected in the sample. For each cluster we report the 
center (RA and Dec) used to build surface brightness profiles, the azimuthal and radial 
positions of 
the cold front and the mean value of $\Delta \alpha$. Bold faced fonts mark clusters
having a merging cold front while italic fonts mark clusters whose merger geometry 
is not clear and the origin of the cold front is not as obvious 
(see \S \ref{sec:occur}).}             
\label{table:CF_sample}      
\centering                          
\begin{tabular}{l c c c c }        
\hline\hline   
Cluster name & center & position angle & jump radius  & $\Delta \alpha$ \\
& & (deg) & (arcsec) & (mean value)\\ 
\hline
Centaurus & 12:48:49.173  -41:18:45.65 & [-60 , 60] & 170 - 210  & 1.60\\
Centaurus & 12:48:49.173  -41:18:45.65 & [90 , 210] & 50 - 100  & 1.76\\
A262& 01:52:46.117 36:09:05.79 & [-110, 0] & 38 - 70  & 1.06 \\
A262& 01:52:46.117 36:09:05.79 & [30, 135] & 40 - 50  & 1.40 \\
Perseus & 03:19:48 41:30:40  & [-60, 0]  &  250 - 300 & 1.72 \\
Perseus & 03:19:48 41:30:40  & [30, 150]  &  140 - 200  & 1.80 \\
A2199 & 16:28:38.193 39:33:02.70 & [-75, -25]  & 25 - 30 & 0.50 \\
A496 & 04:33:38.067 -13:15:40.91 & [-120, -30] &  40 - 55 & 0.79\\
A496 & 04:33:38.067 -13:15:40.91 & [-120, -75] &  180 - 280  & 1.33\\
A496 & 04:33:38.067 -13:15:40.91 & [30, 120] & 80 - 100 &  1.86\\
2A0335+096 & 03:38:40.879  09:58:01.20 & [-120, -30] &  50 - 70 & 1.16 \\
A1644 & 12:57:12.231 -17:24:32.67 & [-180, -90] & 20 - 35 &  1.58\\
A3558 & 13:27:56.989 -31:29:50.00 & [-30, 120] & 80 - 120 & 1.04 \\
A1795 & 13:48:52.879 +26:35:27.80 & [-130 ,-80] & 60 - 70  &  1.14 \\
A2065 & 15:22:29.455 +27:42:23.81 & [-150, -90] & 80 - 100 &  1.26\\
\hline
{\it A576} & 07:21:30.495 +55:45:45.32 & [-120, -60] & 80 - 100 & 0.72 \\
{\it A3562} & 13:33:36.766 -31:40:20.45 & [-150, -60] & 60 - 100  & 1.06\\
\hline 
{\bf A1367} & 11:44:53.5 19:44:19.12 & [-120 -60]  & 350 (~70 from the peak) &  0.84\\
{\bf A754} & 09:09:20.098 -09:40:52.22 & [120,  240]& 80 - 150 & 1.14 \\
{\bf A85} $^*$ & 00:41:42.733 -09:26:33.10 & [0, 90] & 25 - 80 & 1.40  \\
{\bf A3667} & 20:12:41.653 -56:50:52.94 & [-180, -90] & 250 - 280 & 3.68 \\
{\bf A2319}& 19:21:11.097 +43:56:08.00 & [-180, -60] &  100 150 (~160 from the peak) & 1.85\\
{\bf A2256} & 17:02:33.009 +78:38:23.59 & [-135, -90] &  50 - 75 (~100 from the peak) & 2.00\\
{\bf A3266} & 04:31:13.951 -61:27:26.41 & [-90, -30] & 60 100 & 0.87\\
\hline
\hline                                   
\end{tabular}
\end{table*}

\section{Cold fronts occurrence.}
\label{sec:occurr}
\noindent
\subsection{Cold fronts occurrence: a general view}
\label{sub:occ_general}

The exclusion of three unclassified objects (namely A2052, A4059, Hydra A) 
reduces the sample to 42 objects of which 19 host a cold front, 
corresponding to a fraction of 0.45. 
Note that the cold front in the A85 subcluster (dubbed A85$^*$) is included.
The list of the detected cold fronts, with their main properties, is filed in 
Table \ref{table:CF_sample}.

Some clusters (e.g. Centaurus, A496, Perseus, A262) host
more than one cold front. 
This phenomenon is not rare in cool core clusters. {\it Chandra} found multiple cold
fronts in several systems such as A2204, A2029, Ophiuchus 
\citep{Sanders_A2204:2005, Clarke_A2029:2004, AM06,MM_review:2007}. The presence
of multiple cold
fronts in such clusters is likely related to the origin and development of
this phenomenon in cool cores \citep{AM06,MM_review:2007}.
We do not detect any cold front in some objects (i.e. A2204, A2029) , which 
are well-known cold front
systems \citep{Sanders_A2204:2005, Clarke_A2029:2004}.
For these clusters, the cold front is located in very central regions (14\as\
and 30\as\ from the center for A2029 and A2204, respectively). Therefore the 
discontinuity is well-resolved by {\it Chandra}, but it is hard to detect with 
{\it XMM-Newton}.

A last comment concerns the bias in the measure of the occurrence of cold
fronts due to instrumental and observational limits: 
projection effects induce a smoothing on the surface brightness  and
temperature 
profiles and can hide a non-negligible fraction of 
cold fronts. Projection also completely
hides cold fronts having an inclination larger than about 30\deg\ with respect to
the plane of the sky.
In addition, resolution limitations prevent the detection of cold fronts lying
in the very central regions or cold fronts in distant clusters. Moreover, our detection 
algorithm may fail to detect some cold fronts with angular extension $<30$ \deg .
All these effects significantly reduce the capability of detecting cold fronts.
The frequency we find in our sample is  therefore a lower limit of the true 
occurrence.

\subsection{Cold fronts occurrence: relation with redshift} 
\label{sub:redsh}

\begin{figure}
 \centering
\includegraphics[angle=0,width=9.2 truecm]{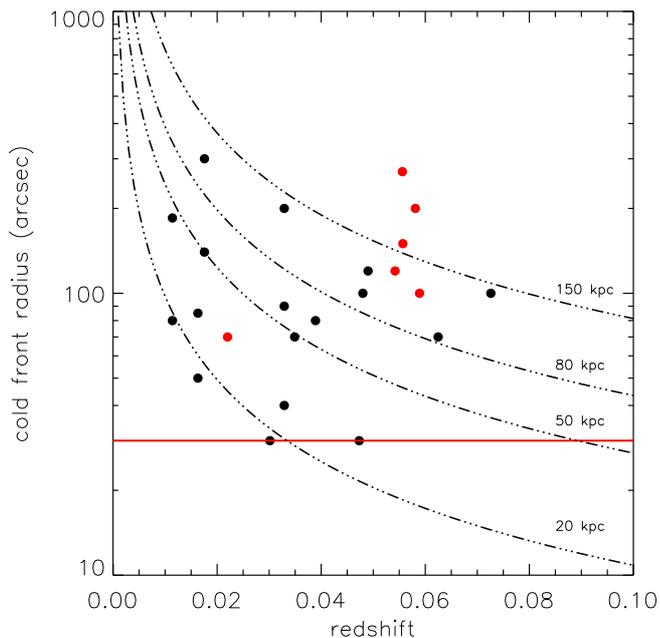}
   \caption{Distances from the cluster center of the cold fronts 
detected in the sample plotted as a function of the redshift
of their hosting systems.
Red points label merging clusters (see Table \ref{table:CF_sample}) 
and black points label the remaining systems. We omitted in this figure A85*  
since the cold front lies in a subclump. Dot dashed
lines plot fixed physical distances at the various redshifts. Red solid line marks the 
{\it XMM-Newton} resolution limit at 30\as .}
   \label{fig:rcf-z}%

\end{figure}

In our sample, no cold fronts are detected in systems at redshifts greater than about $0.075$.
We already remarked that
A2204 and A2029 have been classified as cold front clusters from {\it Chandra} data
analysis but we fail in detecting their discontinuities
because they lie at small distances ($ \simlt 30$\as )
from the X-ray peak, under the {\it XMM-Newton} resolution. For both A2204 and A2029 $z > 0.075$.
This suggests that 
the lack of detection of cold fronts at the highest redshifts of the sample
is most likely related to a resolution limit rather than to a real evolution.
This effect is clearly shown in Fig. \ref{fig:rcf-z} where we plot 
the distances (in arcsec) from the cluster center of all the cold fronts
detected in our sample 
as a function of the redshift of the systems they belong to.
Red points label merging clusters (see Table \ref{table:CF_sample} and \S \ref{sub:merging})
and black points label the remaining systems.
Dashed dotted lines plot fixed physical distances  (20, 50, 80, 150 kpc) at the 
various redshifts.
From this figure, it is evident that cold fronts lying at $\sim$ 20-80 kpc 
from the cluster center are 
observed only in nearby systems ($ z \simlt 0.05$).
Moving towards higher redshifts where these physical distances progressively fall 
below the resolution limit (30\as, red solid line in the figure) cold fronts cannot 
be detected anymore.
For $z > 0.05$, we have found only cold fronts at distances $ \simgt 80-100$ kpc from the 
cluster center, with a 
prominent presence of merging systems whose cold fronts are generally located at large 
distances from the core (see \S \ref{sub:merging}).

On the basis of the analysis of Fig. \ref{fig:rcf-z}, we decided to apply a 
further selection on our sample, namely we cut the maximum redshift at $ z= 0.075$
(i.e. the redshift where we stop detecting cold fronts). 
The resulting sample is reduced to 32 objects 
with a cold front occurrence of $59\%$. We note that the sample may 
be biased against clusters having cold fronts at small distances from the center, inducing 
an underestimation of the cold front occurrence we measure.
However, in the case of cold fronts lying at distances $r \simgt 40$ kpc the sample can be 
considered, to a first approximation unaffected by a redshift bias.
%

\subsection{Occurrence and origin of cold fronts.}
\label{sec:occur}
In this section, we investigate what discriminates clusters without cold fronts from 
clusters hosting one or more.

\subsubsection{Merger cold fronts}
\label{sub:merging}
%

We start by focussing our attention on merger cold fronts.
Some of our systems are well known merging clusters. The morphology of these
systems is generally complex and no unique center can be identified in their surface 
brightness maps. In many of these systems \citep[e.g. A3667][]{Vikhlinin1:2001,
Vikhlinin2:2001, Vikhlinin:2002,Briel:2004}
the merger process is occurring close 
to the plane of the sky and the geometry of the event is clear. 
The origin and the evolution of cold fronts in these systems is most likely related 
to the merger process. More specifically, the motion of a cold dense core of a subsystem 
which moves in the atmosphere of the main cluster during a merger event induces the 
formation of a cold front feature. Typically, the subcluster is stripped of its 
outermost gas and the ram pressure exerted on the surviving dense cloud by the less dense
surrounding gas produces the contact discontinuity between the two subsystems, 
generating the cold front \citep{MM_review:2007}.

In other objects, such as A3562 and  A576, where the X-ray merger geometry 
is not as clear, the nature of the cold fronts we observe is not 
as obvious.
A3562 is a cluster lying in the core of the Shapley supercluster, one of the largest 
mass concentrations in the local Universe.
The presence of a radio halo in this cluster \citep{Giacintucci:2005} provides an indication  
of a merger activity, since radio halos have been found only in interacting systems.
Evidences of interaction come also from {\it Beppo}-SAX data \citep{Bardelli:2002}.
Using {\it XMM-Newton} data, \citet{Finoguenov:2004} suggest that the SC 1329-313
group southwest of A3562 has passed to the north of A3562 and the cluster 
core is likely oscillating in response to the passage of the group.
A576 is another peculiar system whose discontinuity has been 
observed also with {\it Chandra} by \citet{Kempner:2004} who
propose that the core of the cluster is the remnant of a merging subcluster.
This picture was also suggested by \citet{Mohr:1996} from an analysis 
of the galaxy population. Recently, 
\citet{Dupke:2007}  found that the system
is consistent with a line-of-sight merger.
According to this picture, the cold front we find in A576 is likely a 
merger cold front.

The clusters of our sample hosting merger cold fronts 
are listed in the last part of Table \ref{table:CF_sample} and marked 
with a bold-faced font.
We include in this class A85$^*$, the A85 subclump falling on the main structure.
A576 and A3562, which are merging clusters where the cold front origin is not as 
readily associated to the merger event, are  placed in a separate category 
and marked with 
an italic font in Table \ref{table:CF_sample}.

\subsection{Non-merger cold fronts: the entropy profile}
\label{sub:non-merging}

In this subsection we investigate what discriminates clusters without cold fronts 
from clusters hosting at least one, once we exclude the clusters having a merger cold front 
(see above \S\ \ref{sub:merging}) and clusters where the X-ray merger geometry is not clear.
We focus on the remaining subsample (23 clusters) which includes
both clusters undergoing a merger event which is not lying in the plane of the sky and 
clusters which do not present any sign of merging processes. 
In this subsample only 10 out of the 23 clusters host a cold front 
(main properties are listed
in the first part of Table \ref{table:CF_sample}).
We try to understand what determines the presence of cold fronts in these systems, 
studying the radial entropy profile for each of these clusters.
As is conventional in X-ray astronomy, we quantify the entropy 
using the adiabatic constant $K = kT n_e^{-2/3}$ ($T$  and $n_e$ are the 
gas temperature and density respectively and $k$ is a constant) 
following \citet{Voit_entropy:2005}.

The specific 
entropy $s$ is related to $K$ through the relation $s \propto {\rm ln} K$. 
We will refer to $K$ as ``entropy'' throughout the paper.
To obtain the entropy profiles, 
we derive the radial profiles of the electron density $n_e$ and the temperature $T$, 
by deprojecting 
the observed surface brightness and temperature, under the 
assumption of spherical symmetry. The projected temperature and the surface brightness 
have been derived through a spectral analysis of the 
clusters using concentric annuli \citep[see][for details]{Rossetti:2010}.
To perform deprojection, we adopted the 
procedure described in \citet{Ghizzardi_M87:2004}.

\begin{figure}
 
  \centering
   \includegraphics[angle=0,width=9.3 truecm]
    {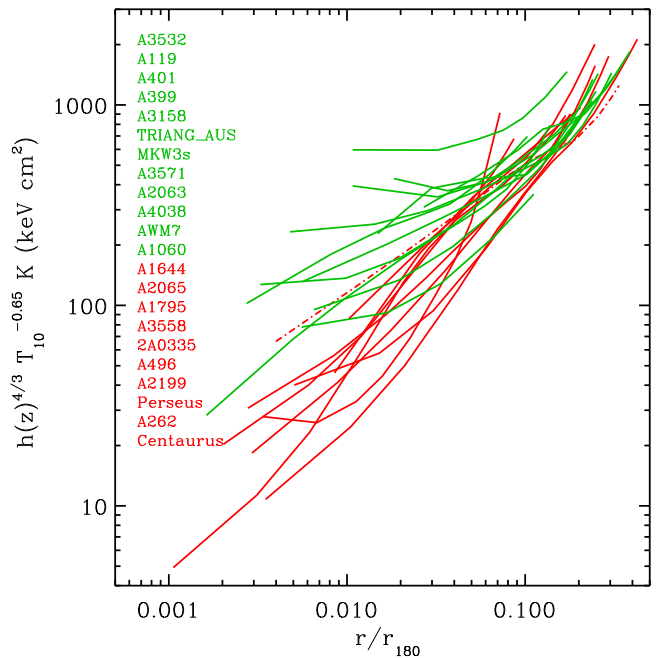}
   \caption{Scaled entropy profiles for the subsample described in \S \ref{sub:non-merging}. 
    Each curve is a 
    locally weighted fit to the data points (see text for details) to reduce 
    the scatter. Red solid lines are the profiles of clusters hosting cold fronts, 
    while green solid lines are the profiles of clusters without cold fronts. 
    The red dot dashed curve is the profile for A3558. 
      Cygnus A is omitted (see text for details).}
   \label{fig:ent_prof_low}%

\end{figure}

In Fig. \ref{fig:ent_prof_low}, we plot all the derived entropy profiles. 
Radii are scaled to $r_{180}$,
the radius within which the mean density is 180 times the critical density\footnote{ $M_{180}=180\rho_c(z)(4\pi/3)r_{180}^3$, where $\rho_c(z)=h^2(z)3H_0^2/8\pi G$ and $h^2(z)=\Omega_{\rm{m}}(1+z)^3+\Omega_{\Lambda}$ }. The values of 
$r_{180}$ have been derived using its 
relationship with the cluster mean temperature as in \citet{Arnaud_r180:2005} 
\citep[see also][]{Leccardi_temp:2008}.
The entropy is scaled using the empirical entropy scaling law $K \propto h(z)^{-4/3} 
T_{10}^{0.65}$,   
$h^2(z) = \Omega_{\rm{m}}(1 + z)^3 + \Omega_\Lambda$ \citep{Pratt_entropy:2006, 
Ponman_entropy_sc:2003};
$T_{10}$ is the mean temperature of each cluster in units of 10 keV, as in 
\citet{Pratt_entropy:2006}.
The curves plotted in Fig. \ref{fig:ent_prof_low} are actually a locally weighted 
regression 
\citep[LOWESS regression, see][]{Sanderson_lowess:2006,Sanderson_lowess:2005} in the 
log-log space to reduce the scatter and provide a better view of the profiles behavior.
Clusters hosting a cold front (hereafter CF clusters) are denoted with red lines
while clusters without cold 
fronts  (hereafter NCF) are denoted with green lines.
We marked as particular case
A3558 (red dot dashed curve). 
The behavior of this cluster will be 
discussed in Sec. \ref{sec:disc}.

The cluster Cygnus A has been discarded here because of the strong contamination by the 
central AGN: the 
presence of the two hot spots invalidates the assumption of spherical symmetry and does 
not allow to deproject the temperature and surface brightness radial profiles.

Fig. \ref{fig:ent_prof_low} shows that all the profiles have a similar trend 
at large radii ($r \simgt 0.08 r_{180}$).
Moving towards the innermost regions, the profiles spread out and we observe a 
large scatter. More precisely, the NCF clusters (green solid curves) typically
have central entropies  higher than CF clusters (red solid curves). Moreover, systems 
hosting cold fronts seem to have a steeper profile than clusters without cold fronts. 

\begin{figure}
 
  \centering
   \includegraphics[angle=0,width=9.5 truecm]
    {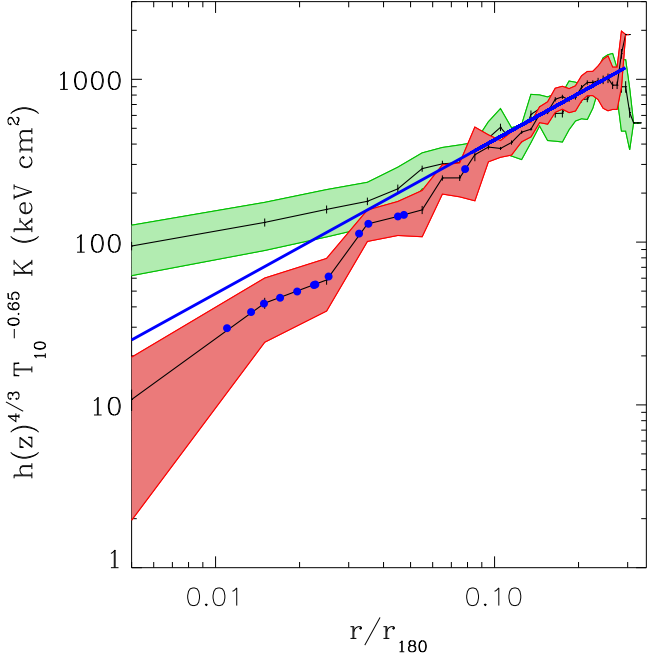}
   \caption{Thin lines are the mean scaled entropy profiles for clusters 
    hosting cold fronts (red area) and 
    clusters without cold fronts (green area). The shaded areas represent the 
    standard deviation from the mean profiles. The thick blue line is the power law 
    obtained fitting all the profiles in the radial range [0.08-0.3]$r_{180}$. 
    Blue points are the distances of the detected cold fronts (excluding A3558) from the 
    cluster center.}
   \label{fig:entropy_shade}%

\end{figure}

In Fig. \ref{fig:entropy_shade} we plot the two averaged profiles (thin solid curves)
for CF and NCF clusters. The shaded areas represent the standard deviation from 
the mean profiles.
As in Fig. \ref{fig:ent_prof_low}, color codes label
CF clusters  (red area) and NCF clusters (green area).
We find that the mean profiles are similar at large radii. 
Fitting all the entropy profiles with a power law in the radial 
range [0.08 - 0.3]$r_{180}$, we find a slope $\alpha=0.95 \pm 0.01$. 
For comparison,  \citet{Pratt_entropy:2006} found
a slope $\alpha = 1.14 \pm 0.06$, while, \citet{Pratt_Arnaud:2005} find 
$\alpha = 0.94\pm 0.14$ and \citet{Piffaretti:2005} find $\alpha = 0.95 \pm 0.02$.
Restricting to $[0.1-0.3] r_{180}$ we find a slightly steeper power law with 
$\alpha=1.08 \pm 0.02$ in accordance with the theoretical 
value of 1.1 predicted by \citet{Tozzi_Norman:2001} 
\citep[see also][]{Voit_Ponman_entropy:2003,Borgani:2002}.

Moving towards the innermost regions, for $r \simlt 0.08 r_{180}$,
the two mean profiles decouple.
The NCF mean cluster profile exhibits a central entropy excess with respect to the 
outer power law model.
The slope in the $[0.01 - 0.08] r_{180}$ range is $\alpha = 0.64 \pm 0.01$ significantly 
lower than the outer power law slope.
On the contrary, the CF cluster profiles become steeper in the same radial range, 
reaching lower central values. Fitting 
with a power law the CF mean entropy profile in the $[0.01-0.08] r_{180} $ range, 
we find a slope of $1.22 \pm 0.01$, significantly higher than the external power 
law slope. The mean entropy values in the innermost bin 
($r=0.005 r_{180}$) are $94.5 \pm 5.5 $ keV cm$^2$ and $10.8 \pm 1.6$
keV cm$^2$ for NCF and CF clusters respectively.
Excluding the particular case 
A3558 (see \S \ref{sec:disc}) 
does not significantly change results on best fit values.

Since the steepness of the entropy profile is an indicator of the presence 
of a ``cool core'' (e.\,g.\, \citealt{Cavagnolo:2009}), clusters with a steep entropy 
profile likely feature also a temperature decrement and a brightness excess in their 
internal regions. Therefore, one may argue that we do not detect cold fronts in objects 
with a flat entropy profile just because their surface brightness is lower than that of 
clusters where we do detect cold fronts. However, as it can be seen from the profiles 
reported in Appendix,  we detect cold fronts in the centers of cool core objects 
(e.\,g.\, A262) where the  $S\!B$ is high, but also in the outer regions of merging 
clusters where the $S\!B$ is much lower (e.g. A3667, A85$^*$). In Fig.\, \ref{fig:discont}, 
we show an example of a cluster where we do not detect CF, AWM7, which has a  $S\!B$ 
comparable to those of the clusters where we do detect cold fronts. This is the case 
for almost all the clusters of our remaining sample where we do not detect cold fronts.

\section{Discussion}
\label{sec:disc}

The origin of cold fronts in clusters manifestly undergoing a merger event 
can be related 
to the motion of a dense cold cloud of gas within the atmosphere of another subcluster.
Conversely, the presence (or the absence) of these features in the subsample of 
non-merging clusters and clusters where the merger is not close to the plane of the sky 
is not clearly understood.
Fig. \ref{fig:entropy_shade} provides some hints to help 
understand what determines the presence of cold fronts in these systems.
The general picture emerging here 
is that the entropy profile discriminates among the two classes (CF and NCF) of 
clusters.
While at large radii the (scaled) entropy profiles of these clusters  are very similar,
in the innermost regions ($r \simlt 0.08r_{180}$) their behaviors differ.

Our finding of a steep entropy gradient in CF cluster is in agreement with 
theoretical expectations. Indeed,  simulations by \citet{AM06} show 
that cold fronts can rise and develop in the cores of clusters
if the entropy sharply decreases towards the center
(as typically occurs in the center of cool core clusters). 
According to these simulations, cold fronts develop as a consequence of minor
merger events;
during its passage near the center of a cluster, a merging subclump induces 
some disturbance on the low entropy gas of the core and displaces it 
from the center of the potential well.
If the entropy profile is steep, the cool gas 
starts sinking towards the minimum of the gravitational potential, a
sloshing mechanism sets in and cold fronts arise.
If the entropy profile is not steep, 
the entropy contrast is insufficient for the cool gas to flow back and 
for the sloshing mechanism to set in.
In agreement with this picture,
we find that cold fronts form only in regions where the entropy profile sharply decreases.
In Fig. \ref{fig:entropy_shade}, we plot (blue filled circles) the cold fronts 
positions measured for
the non-merging clusters of our sample.  Excluding the outermost cold front 
of A496 (this cluster hosts three cold fronts) which lies at a distance of 
$\sim 0.08 r_{180}$ from the peak, all the cold fronts we detect lie at distances 
smaller than $\sim 0.05 r_{180}$, where entropy profiles steepen, and greater 
than $\sim 0.01r_{180}$ where, in many systems, {\it Chandra} detects a 
flattening \citep{Donahue_entropy:2006}.

\begin{figure}
 
  \centering
   \includegraphics[angle=0,width=9.5 truecm]
    {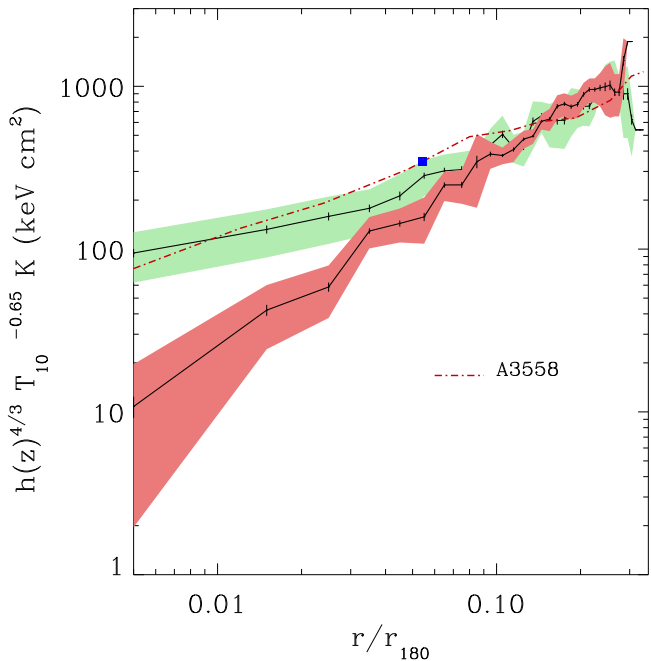}
   \caption{Mean scaled entropy profiles as in Fig. \ref{fig:entropy_shade}.
   Red dot-dashed curve is the profile of A3558. 
   The blue filled square 
   is the A3558 cold front distance from the cluster center. }
   \label{fig:entropy_outliers}%

\end{figure}

The large majority of clusters of the final subsample obey the general rule 
that cold fronts are hosted 
by systems with a steep entropy profile in their centers. However,
as already pointed out, A3558 is a peculiar case: 
although its entropy profile is 
similar to the NCF clusters profiles, it hosts a cold front.
Some comments are needed to understand why this outlier does not follow the 
general behavior.
The cold front for A3558  
(blue square in Fig. \ref{fig:entropy_outliers}) is located at 
larger distance ($ r \sim 0.05 r_{180}$) with respect to all the other cold fronts 
we detect, in a region where a weak entropy gradient, not as sharp as 
in the other CF clusters profiles, is present.

This cluster lies at the center of the 
Shapley supercluster and its special behavior of this cluster is likely related 
to the unique environment in which it is embedded.
To understand the reason why cold fronts can arise in such a system, we refer once more 
to \citet{AM06} simulations.
When the entropy profile of the cluster does not sharply decrease in the center, 
the central cold gas  is easily pushed away from the dark matter peak, at the  merging 
subclump passage. 
Accordingly, the cold front emerges at a large distance from the core. 
However, there is no entropy contrast to 
trigger the sloshing mechanism and this cold front will not develop further.
Cold fronts rising in these systems are short-lived and therefore rare phenomena 
(see Fig.\, 12 in \citealt{AM06}). 
A3558 is embedded in a very unrelaxed environment where merging events 
are frequent, and therefore the probability to form (and to observe) such fronts is higher.
Alternatively, this cold front might be a merger cold front that we failed to recognize 
due to the fact that A3558 cannot be classified easily as a merging cluster.
As discussed in \citet{Rossetti_A3558:2007}, it presents some features similar to those
of cool core clusters and other properties that are more common in merging clusters.

One of the main findings of our paper is that we detect at least one 
cold front in all 
steep entropy gradient clusters in the final subsample. 
\citet{AM06} show that once the sloshing mechanism sets in cold fronts can be 
recognized in all the projection planes, even if they are more prominent on the merger 
plane (see Fig. 19 of their paper). However, the limited resolution of our instruments 
allow us to recognize only the most apparent brightness discontinuity. Indeed, we have 
performed some simulations of cold fronts projection with the {\it XMM-Newton} PSF  and 
we found that cold fronts can only be observed if they lie within some 30\deg\ of the 
plane of the sky. This means that our 100\% detection
rate implies that most steep entropy clusters must host more than one ``prominent'' cold front. 
This abundance of cold fronts suggests that, whatever the triggering mechanism might be, 
it must have a high occurrence rate.  Since the prominent cold fronts that we can detect 
are located on the merger plane,  the detection of one or more cold fronts in all our 
steep entropy systems seems to 
indicate that a sizeable fraction of them are currently experiencing more than one
minor merger. Assuming that, crudely speaking, cold fronts are visible for a timescale 
of about 3 Gyr (this is the case for the dark matter + gas simulation in \citealt{AM06}, 
while for the dark matter only this timescale is longer), our cold front detection rate 
translates into a minimum merger frequency of 
1/3 merger event  per halo per Gyr. If we further assume a minimum mass ratio of 1/10
we can compare our rate with rates expected from cosmological simulations. Using
Fig.8  in a recent paper by \citet{Fakhouri:2008} we find a merger rate of  
$\sim 0.2$ merger per halo per Gyr for mass ratio larger that 1/10. This is 
somewhat smaller than the minimum rate implied by our observed cold front rates
however, given the numerous simplifications we have applied in our calculation, 
we deem it to be in acceptable agreement.

Gas sloshing may provide an important contribution to the 
cooling-heating problem in cool core clusters \citep{ZuHone:2009}.
The sloshing gas typically moves at sub-(or trans-)sonic velocities carrying a 
kinetic energy comparable to the thermal energy but the dissipation of this kinetic 
energy to thermal energy is too slow compared to cooling  \citep{Maxim:2001}. However 
the sloshing mechanism also brings the outer high entropy gas into the core, mixing 
it with the cooling gas and resulting in a heat inflow which can prevent the formation 
of a ``cooling flow'' for periods of time 1-3 Gyr \citep{ZuHone:2009}. If subcluster 
encounters are frequent enough, as it is suggested by our high detection rate, the 
sloshing mechanism can efficiently offset cooling. 
Intriguingly the sloshing mechanism operates preferentially in steep entropy 
profile clusters, i.e. precisely those which require heating to offset the cooling. 
With the coming into operation of the first space-borne micro-calorimeter, 
quite likely the one onboard the ASTRO-H mission \citep{Taka_NEXT:2008}, 
it will be possible to investigate 
gas motions in the direction of the line of sight,
i.e. orthogonally with respect to that of the plane of the sky sampled with cold fronts. 
The combination of the two informations will afford a reliable estimate of 
the motions of the ICM in clusters core and estimate their role in offsetting cooling.

\section{Summary}
\label{sec:summary}

We have performed a systematic search of cold fronts using {\it XMM-Newton} data for a sample 
of 45 objects extracted from the B55 flux limited sample \citep{Edge:B55}.

The main results of our work are the following:

\begin{itemize}

\item{Excluding three unclassified cases, we find that 19 clusters out of 42 
host at least one cold front.}

\item{We do not detect any cold front in systems having redshift greater than about 0.075. 
This is most 
likely related to {\it XMM-Newton} resolution limit. By cutting our sample at
$z = 0.075$, we restrict
our sample to 32 objects with a cold front occurrence of 59\% .}

\item{Cold fronts are easily detected in systems that are manifestly 
undergoing a merger event in (or close to) the plane of the sky.}


\item{Out of the 23 clusters of the remaining subsample (systems 
undergoing a merger 
event which is not lying in the plane of the sky and 
non-merging clusters) 10 objects exhibit a cold front.
For this final subsample, the entropy profile of systems 
hosting cold fronts is found to be steeper than that 
of clusters without them. The difference is observed at radii smaller than about
$0.08 r_{180}$ where all our cold fronts are found.}

\item{Our findings are in agreement with simulation based predictions. As shown by 
\citet{AM06} an entropy gradient is a necessary ingredient to trigger gas sloshing.
}

\item{Since projection effects highly limit the capability of detecting cold fronts, 
the finding that all the clusters with a steep entropy profile host a cold front 
implies that most clusters with a steep entropy profile must have more 
than one cold front. }

\item{Under the assumption that cold fronts in cool core clusters are triggered
by minor mergers, we estimate a minimum of 1/3
events per halo per Gyr, which is somewhat larger than that expected from
cosmological simulations \citep{Fakhouri:2008}.}

\item{Gas sloshing may provide an important contribution to the 
cooling-heating problem in cool core clusters.
A robust assessment of the gas motions associated
to the sloshing phenomenom will become possible with the
coming into operation of the first space borne microcalorimeter.}

\end{itemize}

\begin{acknowledgements}
The authors thank the referee for useful comments
The authors are pleased to acknowledge Sabrina De Grandi and Fabio Gastaldello 
whose suggestions have significantly improved the paper.
\end{acknowledgements}

\bibliographystyle{aa}
\bibliography{biblio}

\begin{appendix}
\section{Surface brightness and temperature profiles}
\begin{figure*}
 \centering
\resizebox{0.8\hsize}{!}{\includegraphics{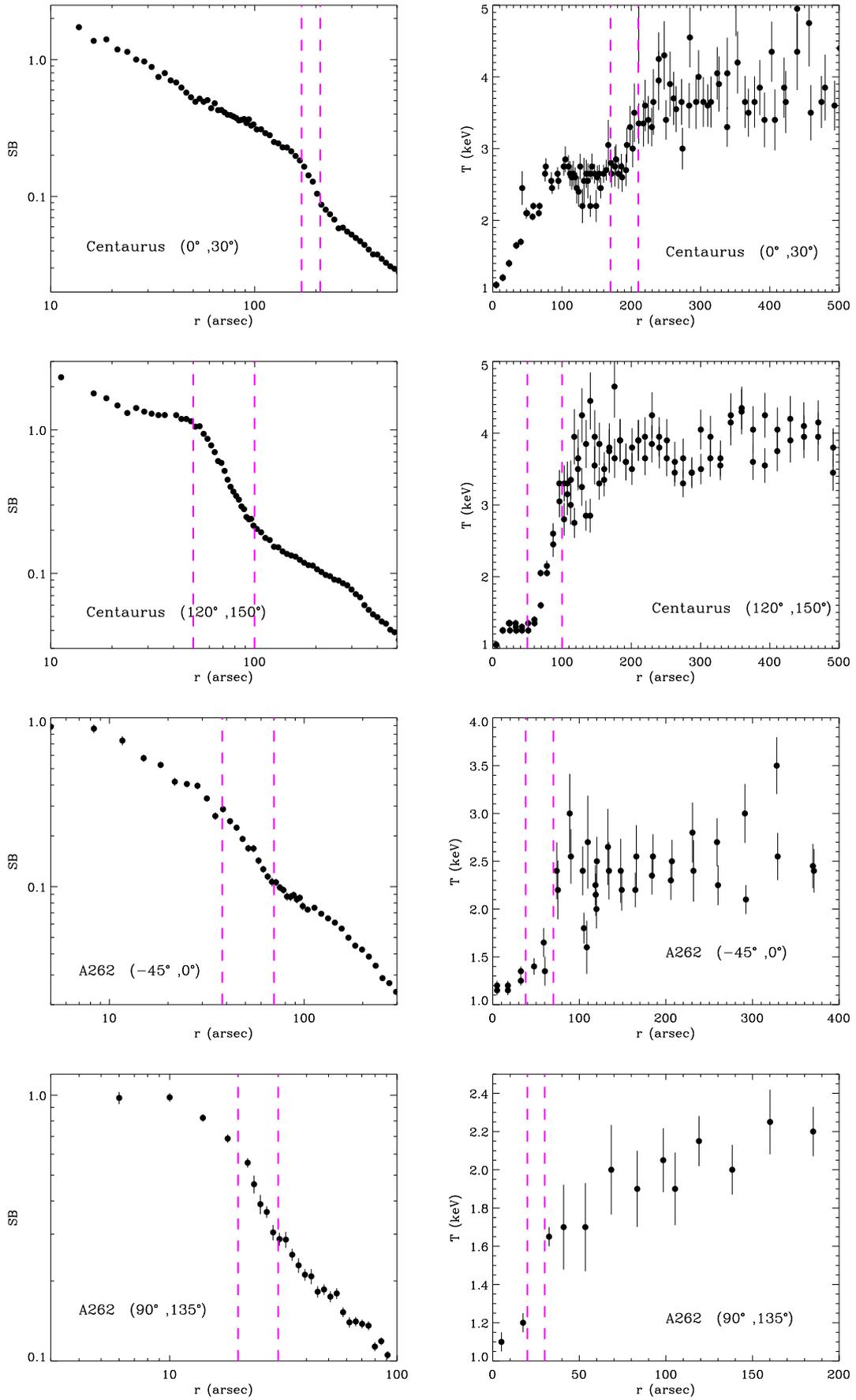}}
   \caption{Surface brightness and temperature profiles for the cold fronts in Centaurus and A262. }
   \label{fig:g1}%
\end{figure*}

\begin{figure*}
 \centering
\resizebox{0.8\hsize}{!}{\includegraphics{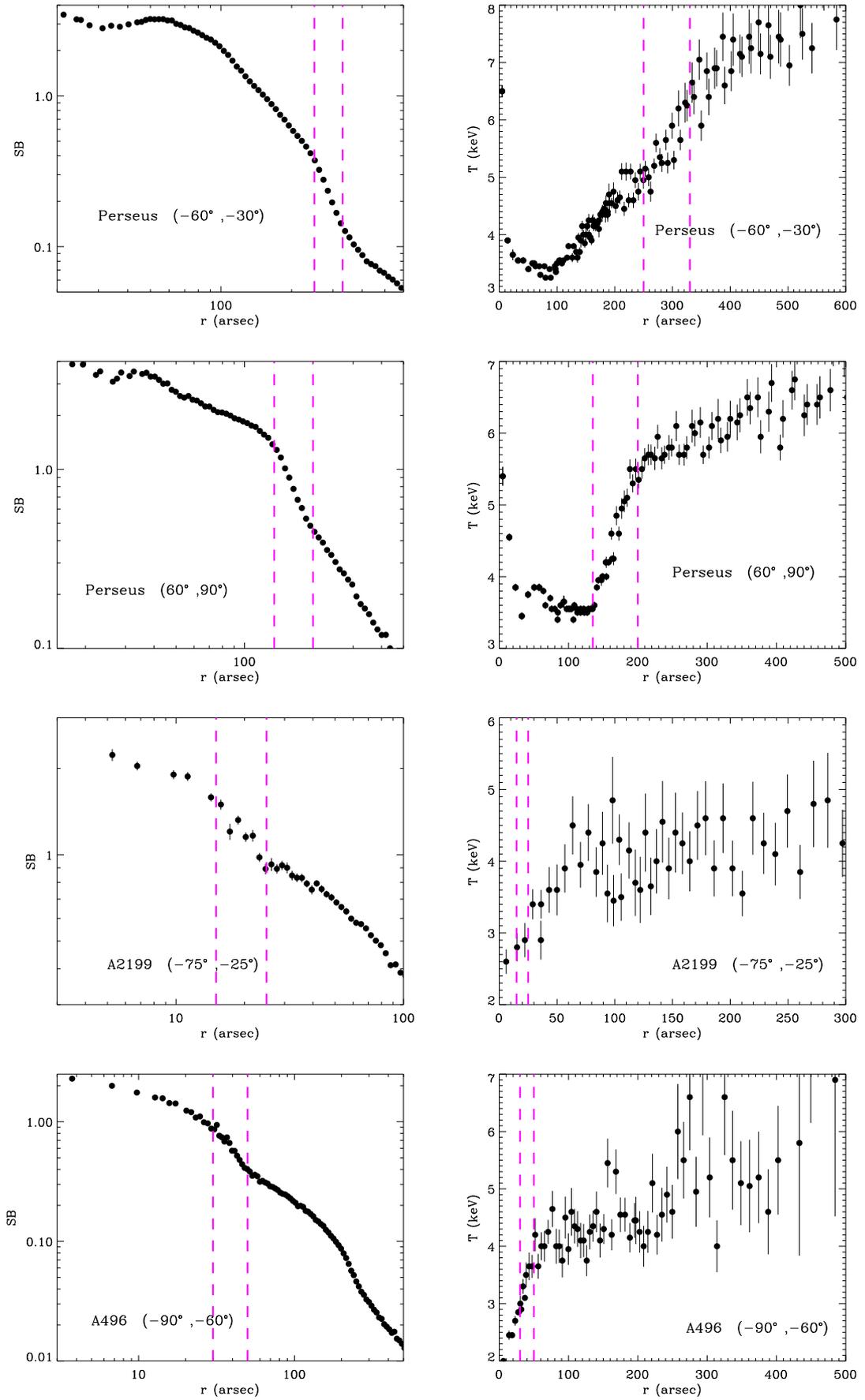}}
   \caption{Surface brightness and temperature profiles for the cold fronts in Perseus, A2199 and A496. }
   \label{fig:g2}%
\end{figure*}

\begin{figure*}
 \centering
\resizebox{0.8\hsize}{!}{\includegraphics{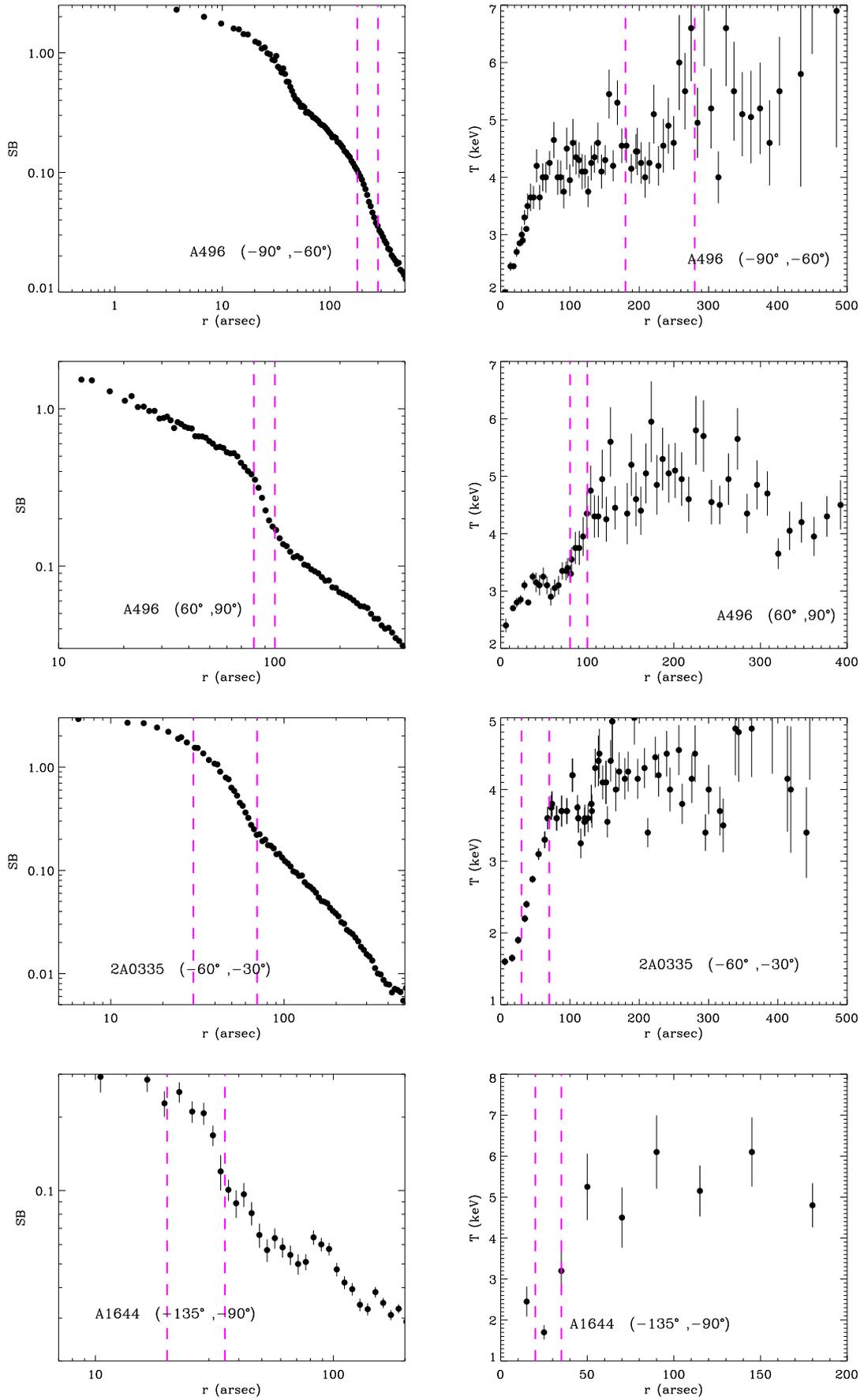}}
   \caption{Surface brightness and temperature profiles for the cold fronts in A496, 2A0335 and A1644. }
   \label{fig:g3}%
\end{figure*}

\begin{figure*}
 \centering
\resizebox{0.8\hsize}{!}{\includegraphics{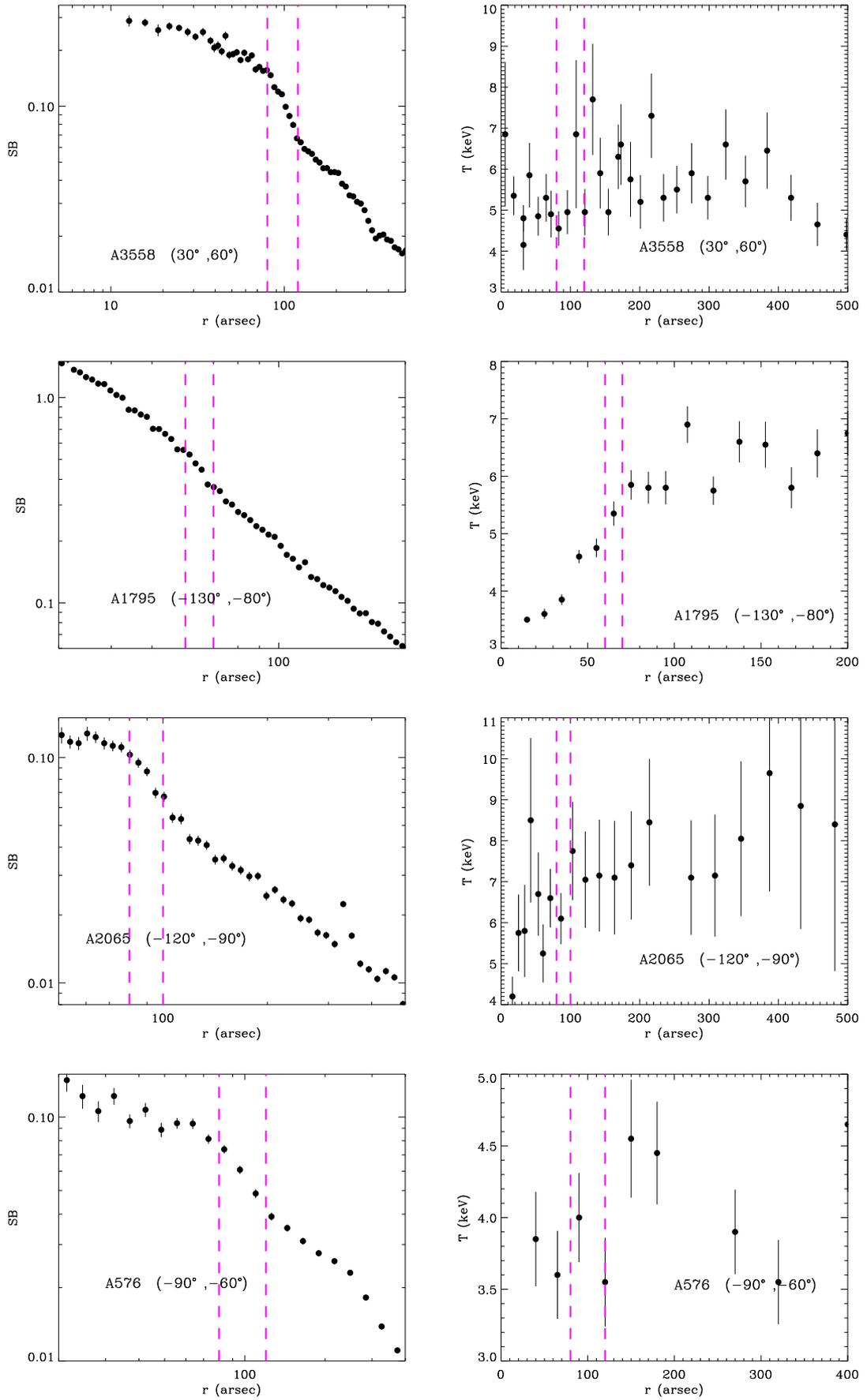}}
   \caption{Surface brightness and temperature profiles for the cold fronts 
in A3558, A1795, A2065 and A576. }
   \label{fig:g4}%
\end{figure*}

\begin{figure*}
 \centering
\resizebox{0.8\hsize}{!}{\includegraphics{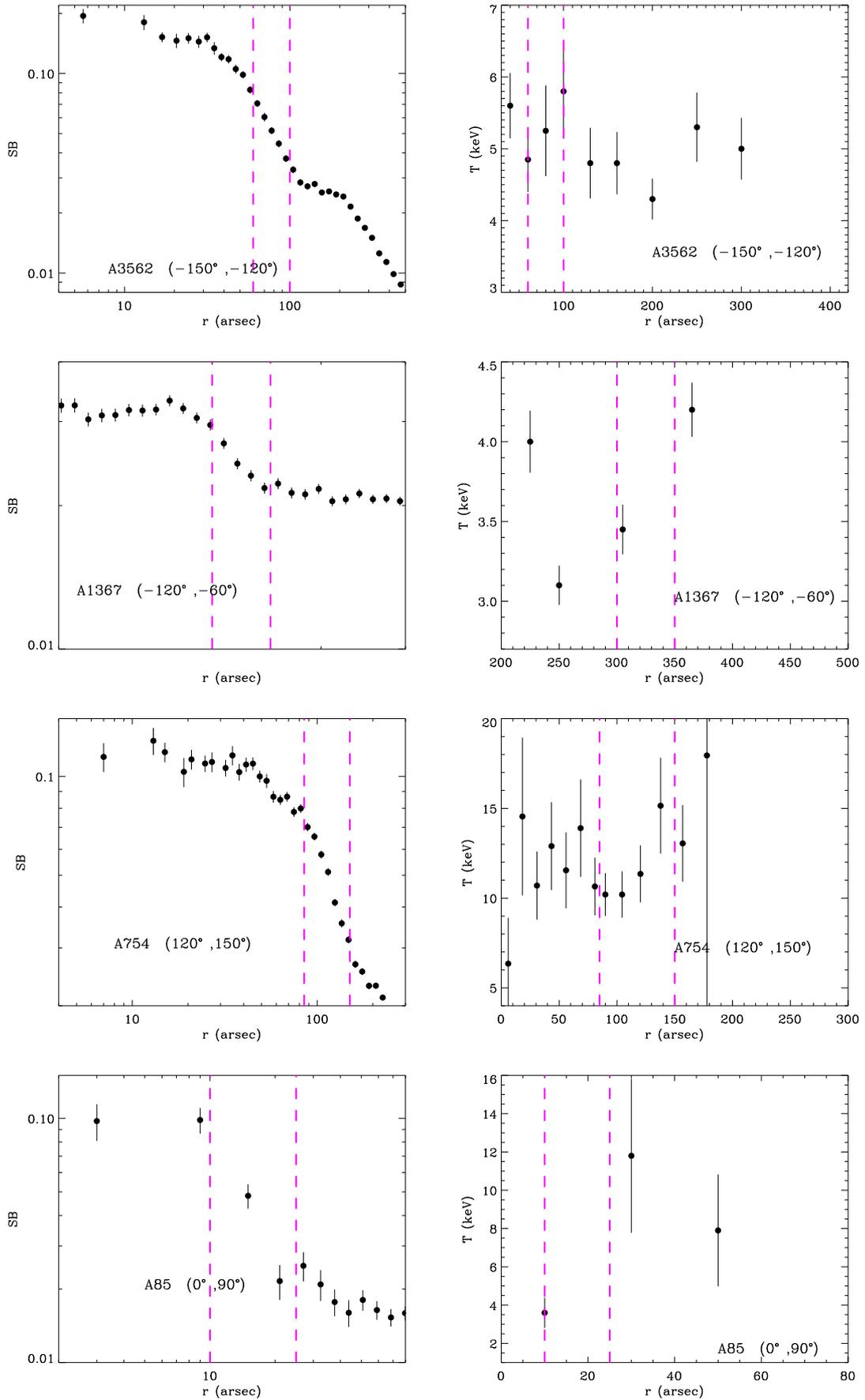}}
   \caption{Surface brightness and temperature profiles for the cold fronts in A3562, A1367, 
A754 and A85$^*$.}
   \label{fig:g5}%
\end{figure*}

\begin{figure*}
 \centering
\resizebox{0.8\hsize}{!}{\includegraphics{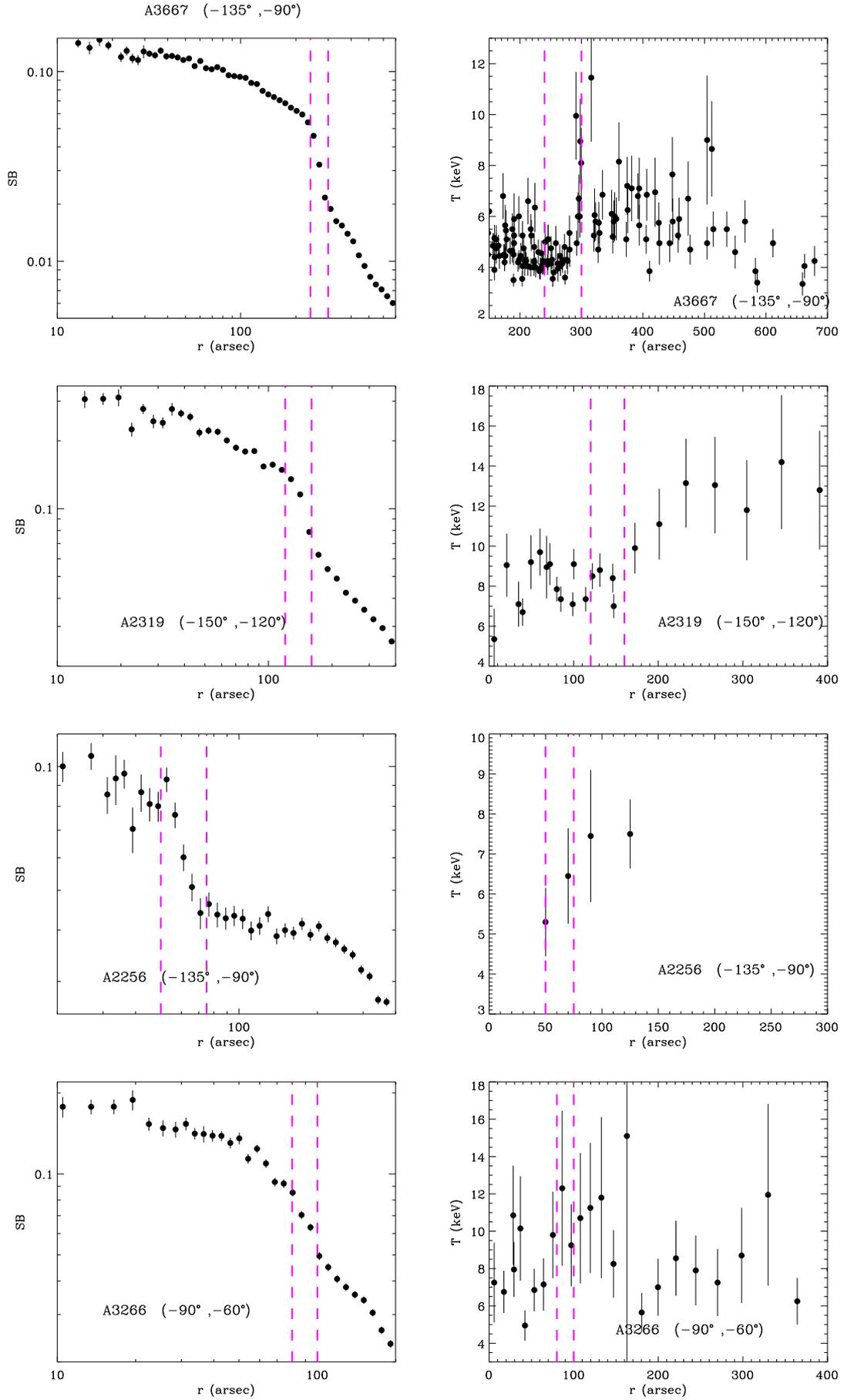}}
   \caption{Surface brightness and temperature profiles for the cold fronts in A3667, A2319, 
A2256 and A3266.}
   \label{fig:g6}%
\end{figure*}

In this Appendix we report the EPIC flux images of the clusters of the sample where 
we detected cold fronts (Full resolution maps are omitted in the astro-ph version of the 
paper and are available at:
http://www.iasf-milano.inaf.it/$\sim$simona/pub/coldfronts/ghizzardi.pdf). 
The figures show the 
flux images in the 0.4-2 keV band, the black arcs indicate the position of the cold 
fronts and the ``X'' symbol the selected center for the extraction of the profiles. 
In Figures \ref{fig:g1}-\ref{fig:g6} we show the surface brightness and temperature 
profiles across the discontinuities in a representative sectors for all the cold 
fronts reported in Table \ref{table:CF_sample}.
For each plot the discontinuity region is marked with vertical red dashed lines.
\end{appendix}

%
%

\end{document}